\PassOptionsToPackage{unicode}{hyperref}
\PassOptionsToPackage{hyphens}{url}
\PassOptionsToPackage{dvipsnames,svgnames,x11names}{xcolor}
\documentclass[12pt]{article}
\usepackage{epsfig,graphicx}
\usepackage{setspace,multirow}
\usepackage{amsmath,amsthm}
\usepackage{amssymb,enumerate}
\usepackage{bbm}
\usepackage[noend]{algorithmic}
\usepackage{setspace,paralist,color}
\usepackage{bm,verbatim,subfigure}
\usepackage{subfigure}
\usepackage{lineno}
\usepackage{multicol}
\usepackage{rotating}
\usepackage{threeparttable}
\usepackage{CJK}
\usepackage[titletoc]{appendix}
\usepackage{titlesec}
\usepackage[round]{natbib}
\usepackage{url}
\usepackage{natbib}
\usepackage{ulem}
\usepackage{iftex}
\usepackage{xr}
\allowdisplaybreaks
\renewcommand{\hat}{\widehat}

\def\be{\begin{eqnarray}}
	\def\ee{\end{eqnarray}}

\def\argmin{\mathrm{argmin}}

\def\pover{\buildrel p\over\longrightarrow}
\newcommand{\E}{{\mathbb{E}}}

\def\n{\nonumber}

\def\sumi{\sum_{i=1}^n}
\def\summ{\sum_{m=1}^M}

\def\train{\mathrm{\rm\tiny train}}
\def\test{\mathrm{\rm\tiny test}}

\def\wh{\widehat}
\def\wt{\widetilde}
\newcommand{\mbN}{{\mathbb{N}}}

\def\bw{{\bf w}}
\def\bX{{\bf X}}
\def\bx{{\bf x}}

\def\calW{{\mathcal{W}}}

\def\bA{{\bf A}}
\def\ba{{\bf a}}

\newcommand{\bg}{{\bm{g}}}

\newcommand{\btheta}{{\bm{\theta}}}

\def\btheta{{\boldsymbol \theta}}

\newcommand{\calD}{{\cal D}}

\newcommand{\calH}{{\cal H}}

\newcommand{\calG}{{\cal G}}

\newcommand{\calY}{{\cal Y}}

\newcommand{\calX}{{\cal X}}

\newcommand{\mbR}{{\mathbb{R}}}

\newcommand{\f}{{\mathbf{f}}}

\newtheorem{Theorem}{\underline{\bf Theorem}}
\newtheorem{Remark}{\underline{\bf Remark}}

\theoremstyle{definition}

\newtheorem{condition}{Condition}

\newtheorem{definition}{\underline{\bf Definition}}

\ifPDFTeX
  \usepackage[T1]{fontenc}
  \usepackage[utf8]{inputenc}
  \usepackage{textcomp} 
\else 
  \usepackage{unicode-math}
  \defaultfontfeatures{Scale=MatchLowercase}
  \defaultfontfeatures[\rmfamily]{Ligatures=TeX,Scale=1}
\fi
\usepackage{lmodern}
\ifPDFTeX\else  
\fi
\IfFileExists{upquote.sty}{\usepackage{upquote}}{}
\IfFileExists{microtype.sty}{
  \usepackage[]{microtype}
  \UseMicrotypeSet[protrusion]{basicmath} 
}{}
\makeatletter
\@ifundefined{KOMAClassName}{
  \IfFileExists{parskip.sty}{%
    \usepackage{parskip}
  }{
    \setlength{\parindent}{0pt}
    \setlength{\parskip}{6pt plus 2pt minus 1pt}}
}{
  \KOMAoptions{parskip=half}}
\makeatother
\usepackage{xcolor}
\makeatletter
\ifx\paragraph\undefined\else
  \let\oldparagraph\paragraph
  \renewcommand{\paragraph}{
    \@ifstar
      \xxxParagraphStar
      \xxxParagraphNoStar
  }
  \newcommand{\xxxParagraphStar}[1]{\oldparagraph*{#1}\mbox{}}
  \newcommand{\xxxParagraphNoStar}[1]{\oldparagraph{#1}\mbox{}}
\fi
\ifx\subparagraph\undefined\else
  \let\oldsubparagraph\subparagraph
  \renewcommand{\subparagraph}{
    \@ifstar
      \xxxSubParagraphStar
      \xxxSubParagraphNoStar
  }
  \newcommand{\xxxSubParagraphStar}[1]{\oldsubparagraph*{#1}\mbox{}}
  \newcommand{\xxxSubParagraphNoStar}[1]{\oldsubparagraph{#1}\mbox{}}
\fi
\makeatother

\usepackage{longtable,booktabs,array}
\usepackage{calc} 
\usepackage{etoolbox}
\makeatletter
\patchcmd\longtable{\par}{\if@noskipsec\mbox{}\fi\par}{}{}
\makeatother
\IfFileExists{footnotehyper.sty}{\usepackage{footnotehyper}}{\usepackage{footnote}}
\makesavenoteenv{longtable}
\usepackage{graphicx}
\makeatletter
\def\maxwidth{\ifdim\Gin@nat@width>\linewidth\linewidth\else\Gin@nat@width\fi}
\def\maxheight{\ifdim\Gin@nat@height>\textheight\textheight\else\Gin@nat@height\fi}
\makeatother
\setkeys{Gin}{width=\maxwidth,height=\maxheight,keepaspectratio}
\makeatletter
\def\fps@figure{htbp}
\makeatother

\makeatletter
\@ifpackageloaded{caption}{}{\usepackage{caption}}
\AtBeginDocument{%
\ifdefined\contentsname
  \renewcommand*\contentsname{Table of contents}
\else
  \newcommand\contentsname{Table of contents}
\fi
\ifdefined\listfigurename
  \renewcommand*\listfigurename{List of Figures}
\else
  \newcommand\listfigurename{List of Figures}
\fi
\ifdefined\listtablename
  \renewcommand*\listtablename{List of Tables}
\else
  \newcommand\listtablename{List of Tables}
\fi
\ifdefined\figurename
  \renewcommand*\figurename{Figure}
\else
  \newcommand\figurename{Figure}
\fi
\ifdefined\tablename
  \renewcommand*\tablename{Table}
\else
  \newcommand\tablename{Table}
\fi
}
\@ifpackageloaded{float}{}{\usepackage{float}}
\floatstyle{ruled}
\@ifundefined{c@chapter}{\newfloat{codelisting}{h}{lop}}{\newfloat{codelisting}{h}{lop}[chapter]}
\floatname{codelisting}{Listing}

\makeatother
\makeatletter
\@ifpackageloaded{caption}{}{\usepackage{caption}}
\@ifpackageloaded{subcaption}{}{\usepackage{subcaption}}
\makeatother

\ifLuaTeX
  \usepackage{selnolig}  
\fi
\usepackage[]{natbib}
\usepackage{bookmark}

\IfFileExists{xurl.sty}{\usepackage{xurl}}{} 
\hypersetup{
  pdftitle={Title},
  pdfauthor={Author 1; Author 2},
  pdfkeywords={3 to 6 keywords, that do not appear in the title},
  colorlinks=true,
  linkcolor={black},
  filecolor={Maroon},
  citecolor={black},
  urlcolor={black},
  pdfcreator={LaTeX via pandoc}}

\newcommand{\anon}{1}

\begin{document}

\def\spacingset#1{\renewcommand{\baselinestretch}%
{#1}\small\normalsize} \spacingset{1}


\if1\anon
{
  \title{\bf Combining pre-trained models via localized model averaging\footnotetext{The authors are listed in alphabetical order and recognized as co-first authors.}}
  \author{Ziwen Gao\hspace{.2cm}\\
    Yau Mathematical Sciences Center, Tsinghua University\\
    Baihua He\hspace{.2cm}\\
    Department of Statistics and Finance, School of Management,\\ University of Science and Technology of China\\
    and \\
    Yuhong Yang\thanks{Corresponding author: yyangsc@tsinghua.edu.cn} \\
    Yau Mathematical Sciences Center, Tsinghua University\\
    Beijing Institute of Mathematical Sciences and Applications}
  \maketitle
} \fi

\if0\anon
{
  \bigskip
  \bigskip
  \bigskip
  \begin{center}
    {\LARGE\bf Combining pre-trained models via localized model averaging}
\end{center}
  \medskip
} \fi

\bigskip
\begin{abstract}
Many pre-trained models (PTMs) are available in modern applications. Because different PTMs are often trained on different datasets, their performances can vary substantially for different new tasks, and the ranking of the candidates may depend heavily on the input. 
Motivated by this, we propose a localized model averaging method with weights modeled as functions of the covariates, making it substantially more versatile than existing model averaging methods. This formulation allows the model averaging procedure to adaptively capture the varying relative advantages of different PTMs across heterogeneous contexts. Specifically, we learn flexible local weights under a general loss framework that accommodates a broad class of prediction tasks. We further establish the asymptotic optimality of the proposed method for both in-sample and out-of-sample risks, as well as the consistency of the estimated weights. Extensive numerical experiments further demonstrate the effectiveness of the proposed method.
\end{abstract}

\noindent%
\vfill

\newpage
\spacingset{1.8} 

\section{Introduction}
In many modern applications, researchers have access to multiple pre-trained models (PTMs) provided by different companies. For example, in image classification tasks, some PTMs (ResNet, DenseNet, GoogLeNet, etc.) are pre-trained on large benchmark datasets such as ImageNet and are subsequently made available for prediction tasks.  
Because different PTMs are trained on different datasets and possibly have different structures, they excel at different tasks and their relative performances may heavily rely on the specific value of the input. In addition, with a large data set, one may split the data and use one part to train a set of learning procedures, and then focus on combining the estimates to obtain a better one in terms of accuracy. 
Motivated by these applications, we advocate the approach of combining different PTMs via localized model averaging to achieve optimal prediction performance on a new task. 

Most existing model averaging methods are essentially global in nature, i.e., the weights are independent of the covariates $\bX$, such as forecast combination \citep{bates1969combination}, Bayesian model averaging \citep{clyde1996prediction,hoeting.madigan.ea:1999}, smoothed information criteria \citep{buckland.burnham.ea:1997,hjort2003frequentist,claeskens.croux.ea:2006}, adaptive weighting \citep{yang:2001,yuan.yang:2005,yu2025adaptively}, and optimal weighting \citep{hansen:2007,zhang.wan.ea:2013,li2022adaboost}. 
Existing theoretical studies on frequentist model averaging have primarily focused on adaptation \citep{yang:2001,yuan.yang:2005},  minimax optimal aggregation \citep{Nemirovski2000Functional,yang2000combining,yang2000mixing,tsybakov2003optimal,catoni2004statistical,yang2004aggregating,leung2006information,Dalalyan2012Sharp,wang2014adaptive}, and asymptotic optimality \citep{hansen:2007,zhang2020parsimonious,peng2022improvability}. In this work, we focus on the asymptotic optimality perspective, which 
includes both in-sample and out-of-sample loss/risk optimality. While most existing studies \citep{WAN:2010,xie:2015,fang2019model,peng2025optimality} focus on asymptotic optimality in terms of in-sample loss/risk, the asymptotic optimality of out-of-sample loss/risk has received much less attention. Two examples are \cite{zhang2023model} and \cite{hu2023optimal}, which deal with parametric or semiparametric cases. In this work, we consider PTMs as candidate models.

Since different companies typically use different training data for their PTMs, they may perform quite differently according to the input $\bX$ in different regions of the covariate space. It is thus desirable to assign weights that depend on $\bX$ rather than using global model averaging (GlobalMA). 
In this paper, we propose a localized model averaging (LocalMA) method to combine different PTMs.
Specifically, the weights are treated as functions of $\bX$ and are estimated via a neural network, where $\bX$ serves as the input and the output is processed through a softmax function. Related to our work, there are two studies that adopt a localization-based approach. 
\cite{pan2006using} propose an input-dependent weighting method, where the weight for each model represents the probability that the model makes a correct prediction given the input. 
\cite{yang2008localized} proposes localized model selection methods and derives their corresponding theoretical properties.

\begin{figure}[htbp]
	\centering
	\subfigure[Compare estimates with true regression function]{
		\includegraphics[width=7.2cm]{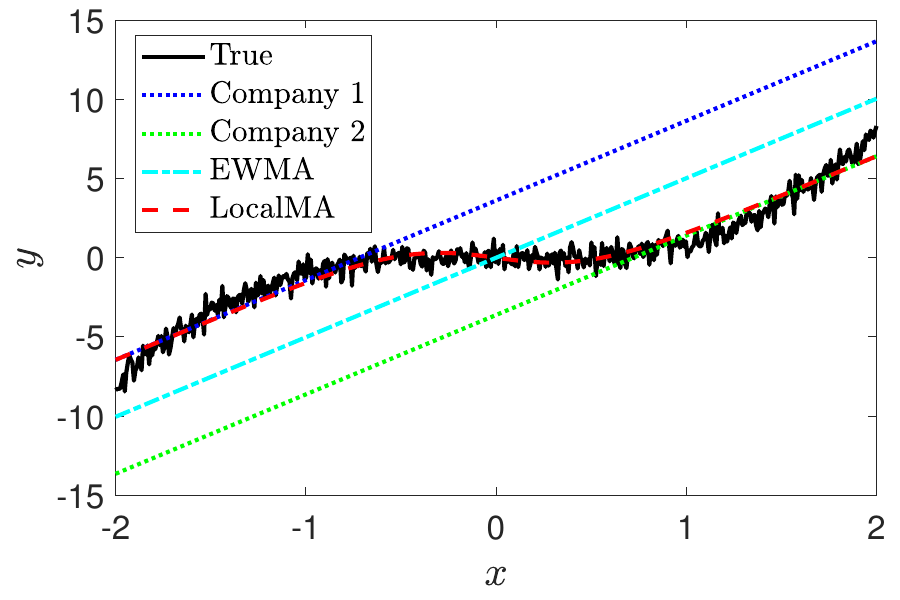}\label{fig:motivationa}
	}
	\subfigure[Weight distribution in the LocalMA.]{
		\includegraphics[width=7.2cm]{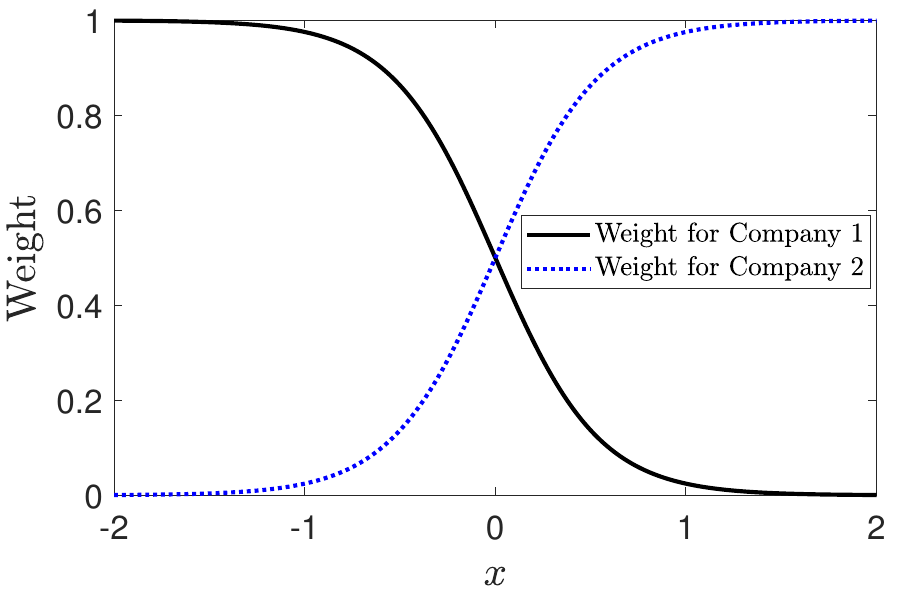}\label{fig:motivationb}
	}
	\caption{A motivating example.}
	\label{fig:motivation}
\end{figure}

We give a simple illustration of the difference between the GlobalMA and LocalMA methods. Consider estimating the regression function $f_0(x)=x^3$ on $[-2,2]$ using $n=200$ observations, where the covariate $x_i$ is uniformly distributed on $[-2, 2]$, $y_i=f_0(x_i)+\varepsilon_i$, and $\varepsilon_i\sim\text{N}(0,0.5^2)$. Suppose Company 1 collects 1000 observations only with $x\in[-2,-0.5]$, and obtains a linear model trained on its data. Company 2 collects 1000 observations only with $x\in[0.5,2]$, also obtaining a fitted linear model. Figure \ref{fig:motivationa} shows the performances of the equal-weight model averaging (EWMA) and the LocalMA method (introduced in Section \ref{sec:modelsetup}). In this case, the EWMA method represents the best performance of GlobalMA. From Figure \ref{fig:motivationa}, we can see that the LocalMA method fits the true regression function well. From Figure \ref{fig:motivationb}, it is clear that in the region where Company 1 or 2 performs better, the LocalMA method assigns higher weights. This example illustrates the advantage of the LocalMA method.

In machine learning, there is a similar idea known as the mixture of experts (MoE) method. The MoE framework proposed by \cite{jacobs1991adaptive} involves a form of model averaging. MoE consists of a set of experts and a gating network, where the gating network dynamically adjusts the weights according to $\bX$ to combine the predictions from multiple experts. 
MoE has been widely applied in large language models \citep{Shazeer2017,fedus2022switch,dai2024deepseekmoe,akbarian2024quadratic,cai2025survey}. 
Our method differs from MoE in two major aspects: in MoE, both the parameters of the experts and those of the gating network are jointly estimated, whereas in our framework, only the parameters in the weighting function need to be estimated. Moreover, while most MoE studies model the gating network as a linear transformation followed by a softmax function \citep{lepikhin2020gshard}, our weighting function applies the softmax to the output of a neural network, making it more general and flexible.

In this background, two questions naturally arise: (1) Given the much more flexible weights in LocalMA, can it achieve asymptotic optimality as enjoyed by GlobalMA? (2) Since the asymptotic optimality of out-of-sample loss/risk is particularly important in practical prediction tasks, can our LocalMA method attain such optimality?

This paper addresses these two questions and makes four main contributions. (1) We propose a novel LocalMA method that captures the varying relative advantages of different PTMs. (2) Our method is developed under a general loss framework and is applicable to a range of common machine learning tasks, such as regression prediction, image classification, and natural language processing. (3) Under certain regularity conditions, we establish the asymptotic optimality of LocalMA for both in-sample and out-of-sample risks, as well as the consistency of the estimated weights. (4) This work is the first to employ a neural network for estimating the weights in optimal model averaging and to develop corresponding theoretical insights. The results from finite sample simulations and real data applications demonstrate the effectiveness of the proposed method. 


The rest of this paper is organized as follows. In Section \ref{sec:modelsetup}, we present the model setup and the procedure for estimating the weights. In Section \ref{sec:theory}, we establish the main theoretical properties of the proposed estimator. The numerical results of the simulation experiments and three datasets are presented in Sections \ref{sec:simulation} and \ref{sec:dataset}, respectively.   Concluding remarks are given in Section \ref{sec:concluding}. Proofs of the theorems appear in the online Supplementary Material.

\section{Problem setup and proposed method}\label{sec:modelsetup}
Suppose we collect $n$ independent and identically distributed (i.i.d.) observations $\mathcal{D}=\{V_i=(\bX_i,Y_i)\}_{i=1}^n$ from the target distribution $\mathbb{P}$, where $\bX_i\in\calX$ is the vector of $p$-dimensional covariates, and $Y_i\in\calY$ is the response for the $i$-th observation, respectively. Define $f_0$ as the true regression function of the target distribution $\mathbb{P}$, i.e., $f_0(\bx)=\E(Y|\bX=\bx)$, where $(\bX,Y)$ is from $\mathbb{P}$. 
In this work, we focus on estimating the regression function $f_0$. The estimators $\hat f_1,\ldots,\hat f_M$ of $f_0$, provided by the $M$ PTMs (which may also be treated as fixed functions), are assumed to be independent of $\mathcal{D}$. 
Let $V^\ast=(\bX^\ast,Y^\ast)$ denote an independent new observation drawn from the distribution $\mathbb{P}$. 
We aim to combine the $M$ PTMs based on $\mathcal{D}$ by LocalMA and predict the response at $\bX^\ast$.

Traditional model averaging \citep{hoeting.madigan.ea:1999,yang:2001,hansen:2007} or GlobalMA assumes that the weights are independent of the covariates:
\be
\wh f_\bw(\bX^\ast)=\summ w_m \wh f_m(\bX^\ast),
\ee
where 
$\bw=(w_1,\ldots,w_M)^\top$ is a constant weight vector belonging to the set $\calW=\{\bw\in[0,1]^{M}:\summ w_m=1\}$. In our problem, it is important to assign weights that depend on  $\bX$, since different PTMs may exhibit different performances in different regions of the covariate space.
We construct the LocalMA prediction as follows:
\be\label{LMAsoft}
\wh f_{\bw_{\bX^\ast}}(\bX^\ast) = \summ w_m(\bX^\ast) \wh f_m(\bX^\ast),
\ee
where $\bw_{\bX^\ast}=(w_1(\bX^\ast),\ldots,w_M(\bX^\ast))^\top$ is an input-dependent weight vector belonging to the set $\calW_{\bX^\ast} =\{ \bw_{\bX^\ast}={\rm softmax}\{\bg(\bX^\ast)\}: \bg\in\mathcal{G}\}$, and $\bg(\bX)=(g_1(\bX),\ldots,g_M(\bX))^\top$ is a function from $\mathbb{R}^p$ to $\mathbb{R}^M$. Here, $\calG$ is a proper function class. For analytical convenience, we assume that each element of the function $\bg\in\calG$ belongs to a H\"{o}lder class in the next section. Note that  $w_m(\bX^\ast)=\exp\{g_m(\bX^\ast)\}/\sum_{j=1}^{M}\exp\{g_j(\bX^\ast)\}$ for $m=1,\ldots,M$. For identifiability, we impose the constraint $g_M = 0$ throughout this paper. 

We use a neural network (NN) to establish the relationship between weights and covariates:
\be\label{w_NN_x}
\bw_{\bX} &=& (w_1(\bX),\ldots,w_M(\bX))^\top\n\\
&=&\sigma_D(\bA_D\sigma_{D-1}(\cdots\sigma_1(\bA_1\bX+\mathbf{b}_1)\cdots)+\mathbf{b}_D)\equiv\sigma_D(\wt\bg(\bX)),
\ee
where 
$\bA_l\in\mathbb{R}^{p_{l+1}\times p_l}$ and $\mathbf{b}_l\in\mathbb{R}^{p_{l+1}}$ are parameters in NN for $l=1,\ldots,D$, $p_1=p$ is the dimension of the input $\bX$, $p_{l+1}(l=1,\ldots,D-1)$ are the number of neurons in the $l$-th hidden layer,  $p_{D+1}=M$ is the dimension of the output layer, $\sigma_l(\cdot)$ $(l=1,\ldots,D-1)$ are the ReLU activation functions, and we consider $\sigma_D(\cdot)$ is the softmax function. We assume that each element of $\wt\bg(\bX)=(\wt g_1(\bX),\ldots,\wt g_M(\bX))^\top$ satisfies $\|\wt g_m\|_\infty\leq B_g$ for some $0<B_g<\infty$, where $\|f\|_\infty$ is the sup-norm of a function $f$. 
The estimates of the parameters $\{\bA_l,\mathbf{b}_l\}_{l=1}^D$ are obtained by minimizing the loss 
\be\label{LossNN}
L_{n}(\bw_{\mathcal{X}},\wh\f) = \frac{1}{n}\sum_{i=1}^n \ell\left\{Y_i, \summ w_m(\bX_i)\wh f_m(\bX_i)\right\},
\ee
where $\bw_{\mathcal{X}}=(\bw_{\bX_1},\ldots,\bw_{\bX_n})$, $\wh\f=(\wh f_1,\ldots,\wh f_M)$, and $\ell(\cdot)$ is a loss function. The loss $\ell$ is chosen properly for the specific regression problem. For example, for continuous regression, $\ell$ can be taken as the standard square loss, and for classification, the cross entropy is natural. In the case of generalized linear model (GLM) regression framework (e.g., Poisson/Gamma regression, or Tweedie regression), the Kullback–Leibler (KL) loss is mathematically convenient.

The resultant estimates of the parameters $\{\bA_l,\mathbf{b}_l\}_{l=1}^D$ are:
\be
\{\wh{\bA}_l,\wh{\mathbf{b}}_l\}_{l=1}^D = \argmin_{\{\bA_l,\mathbf{b}_l\}_{l=1}^D}\text{ }L_{n}(\bw_{\mathcal{X}},\wh\f).
\ee
Then, by plugging $\{\wh\bA_l,\wh{\mathbf{b}}_l\}_{l=1}^D$ into Equation \eqref{w_NN_x}, we obtain the estimate of the weights: 
\be\label{hatwab}
\wh\bw_{\bX_i} = (\wh w_1(\bX_i),\ldots,\wh w_M(\bX_i))^\top = \sigma_D(\wh\bA_D\sigma_{D-1}(\cdots\sigma_1(\wh\bA_1\bX_i+\wh{\mathbf{b}}_1)\cdots)+\wh{\mathbf{b}}_D)
\ee
for $i=1,\ldots,n$ and $\wh\bw_{\mathcal{X}}=(\wh\bw_{\bX_1},\ldots,\wh\bw_{\bX_n})$. Finally, the LocalMA prediction at an input $\bX^\ast$ is 
\be
\wh f_{\wh\bw_{\bX^\ast}}(\bX^\ast) = \summ \wh w_m(\bX^\ast)\wh f_m(\bX^\ast),
\ee
where $\wh\bw_{\bX^\ast} = (\wh w_1(\bX^\ast),\ldots,\wh w_M(\bX^\ast))^\top = \sigma_D(\wh\bA_D\sigma_{D-1}(\cdots\sigma_1(\wh\bA_1\bX^\ast+\wh{\mathbf{b}}_1)\cdots)+\hat{\mathbf{b}}_D)$.

\section{Theoretical Results}\label{sec:theory}
In this section, we present the asymptotic properties of the LocalMA estimator. 
Based on the loss function $L_{n}(\bw_{\mathcal{X}},\wh\f)$, we can define the in-sample risk function as 
$
R_{\rm in}^0(\bw_{\mathcal{X}},\wh\f) =n^{-1} \sum_{i=1}^n \mathbb{E}_{Y|\bX}[\ell\{f_0(\bX_i),\wh f_{\bw_{\bX_i}}(\bX_i)\}]\overset{(a)}{=}n^{-1} \sum_{i=1}^n \ell\{f_0(\bX_i),\wh f_{\bw_{\bX_i}}(\bX_i)\}\n
$,
where $(a)$ is from the independence between $\wh f_m$ and $\calD$. Since $\ell$ has no specific form, we need to make an assumption about its behavior, i.e., 
\be\label{assumprisk}
\frac{1}{n}\sumi\ell\left\{f_0(\bX_i),\wh f_{\bw_{\bX_i}}(\bX_i)\right\}=\frac{1}{n}\sumi\mathbb{E}_{Y|\bX}\left[\ell\left\{Y_i,\wh f_{\bw_{\bX_i}}(\bX_i)\right\}\right]-\delta_{n,\ell}, 
\ee
where $\delta_{n,\ell}$ is a quantity that may depend on $\ell$ but not on the $Y_i$ values.
Under this assumption, we have $$R^0_{\rm in}(\bw_{\mathcal{X}},\wh\f)=\frac{1}{n}\sumi\mathbb{E}_{Y|\bX}\left[\ell\left\{Y_i,\wh f_{\bw_{\bX_i}}(\bX_i)\right\}\right]-\delta_{n,\ell}\equiv R_{\rm in}(\bw_{\mathcal{X}},\wh\f)-\delta_{n,\ell}.$$ Similary, define the out-of-sample risk as
\be\label{assumpriskout}
R_{\rm out}^0(\bw_{\bX^\ast})&=&\mathbb{E}_{V^\ast}\left[\ell\left\{f_0(\bX^\ast),\wh{f}_{\bw_{\bX^\ast}}(\bX^\ast)\right\}\right] \n\\
&=&\mathbb{E}_{V^\ast}\left[\ell\left\{Y^\ast,\wh{f}_{\bw_{\bX^\ast}}(\bX^\ast)\right\}\right]-\delta^\ast_\ell\n\\
&\equiv& R_{\rm out}(\bw_{\bX^\ast})-\delta^\ast_\ell,
\ee
where $\E_{V^\ast}$ is the expectation taken over $V^\ast=(\bX^\ast,Y^\ast)$ and $\delta^\ast_\ell$ is a quantity that may depend on $\ell$ but not on the $Y^\ast$ values.
\begin{Remark}
	It is readily seen that in \eqref{assumprisk} and \eqref{assumpriskout}, $\delta_{n,\ell}$ and $\delta^\ast_\ell$ only serve as the normalization quantities, and many loss functions take this form. If $\ell$ is the KL loss under the GLM, then $\delta_{n,\ell}=\delta^\ast_\ell=0$ (see Supplementary Material for details); and if $\ell$ is the square loss for continuous regression, then $\delta_{n,\ell}=n^{-1}\sum_{i=1}^n\mathbb{E}_{Y|\bX}\left[\ell\left\{Y_i,f_{0}(\bX_i)\right\}\right]$ and $\delta^\ast_\ell=\E_{V^\ast}\left[\ell\left\{Y^\ast,f_{0}(\bX^\ast)\right\}\right]$. 
\end{Remark}
We state some regularity conditions for establishing asymptotic properties. Define $[n]=\{1,\ldots,n\}$ for any positive integer $n$.  
\begin{condition}\label{convF}
	There exists a sequence $a_N\to\infty$ and a positive constant \( c \) such that uniformly for each \( m \in [M] \),
	$$\max_{i\in[n]} {\E}^{1/4}\left[\left|a_N\left\{\wh{f}_{m}(\bX_i)- f_{m}^*(\bX_i)\right\}\right|^4\right] \leq c\;
	\mbox{and}\; \max_{i\in[n]}  \left|f_{m}^*(\bX_i)\right| \leq c.$$
\end{condition}
In Condition \ref{convF}, $f^\ast_{m}(\bX_i)$ is the limiting value of $\wh{f}_{m}(\bX_i)$, closely related to the notion of pseudo-true value (\citealt{white1982maximum}; \citealt{gospodinov2021generalized}; \citealt{lv2014model}). The sequence $a_N\rightarrow\infty$ yields an upper bound on the convergence rates of the estimators $\wh f_1,\ldots,\wh f_M$.
Let $f^{\ast}_{\bw_{\bX}}(\bX)=\summ w_m(\bX)f^\ast_m(\bX)$. 
\begin{condition}\label{losssw}
	There exist two pairs of fixed positive real numbers $(B_\ell,\gamma)$ and $(\wt B_\ell,\wt\gamma)$ such that the random variable $\ell\{Y, f^\ast_{\bw_{\bX}}(\bX)\}|\bX$ is $(B_\ell,\gamma)$-sub-Weibull and $\ell\{Y, \wh f_{\bw_{\bX}}(\bX)\}|\wh\f$ is $(\wt B_\ell,\wt\gamma)$-sub-Weibull.
\end{condition}
The definition of sub-Weibull can be found in the Supplementary Material. By leveraging the general probabilistic tools concerning sub-Weibull random variables and their concentration properties, we are able to handle more general, possibly unbounded loss functions. 
When $\gamma=1/2$, $\ell$ is sub-Gaussian; when $\gamma=1$, $\ell$ is sub-exponential; and $\gamma=0$ indicates that $\ell$ is uniformly bounded. In addition, we provide several examples for the cases where the loss function $\ell$ is the squared loss or the cross-entropy loss, illustrating how to establish that $\ell\{Y, f^\ast_{\bw_{\bX}}(\bX)\}|\bX$ and $\ell\{Y, \wh f_{\bw_{\bX}}(\bX)\}|\wh\f$ are sub-Weibull. The detailed derivations can be found in the Supplementary Material. 


\begin{condition}\label{lip}
	The loss function is Lipschitz continuous, that is, there exists a positive constant \( c_1\) such that for any uniformly bounded functions \( f_i \) and \( f_i' \), 
	$$\left|\ell(Y_i,f_i) -\ell(Y_i,f_i')\right| \leq H(V_i) \left|f_i - f_i'\right|,$$
	where $V_i=(\bX_i,Y_i)$ and $\max_{i\in[n]} \mathbb{E}^{1/4}\left|H(V_i)\right|^4 \leq c_1$.
\end{condition}

Condition \ref{lip} regulates the behavior of the loss function. This implies that $\ell$ is stochastically equicontinuous with respect to $f_i$. A similar form of this condition can be found in Assumption 2 of \cite{yu2025unified}. 
\begin{definition}
	Let $\kappa=\alpha+\nu>0$, $\nu\in(0,1]$ and $\alpha=\lfloor\kappa\rfloor\in \mbN_0$, where $\lfloor\kappa\rfloor$ denotes the largest integer strictly smaller than $\kappa$ and $\mbN_0$ denotes the set of nonnegative integers. For a finite constant $c_0>0$, define the H\"{o}lder class $\calH^{\kappa}(\calX, c_0)$ as
	\begin{align*}
		\calH^{\kappa}(\calX, c_0)=\left\{g:\calX\rightarrow \mbR: \max\limits_{\|\ba\|_1\leq \alpha} \|\partial^\ba g\|_{\infty}\leq c_0, \max\limits_{\|\ba\|_1=\alpha}\sup_{\bx\neq \bx^\prime} \frac{|\partial^\ba g(\bx)-\partial^\ba g(\bx^\prime)|}{\|\bx-\bx^\prime\|_2^{\nu}}\leq c_0\right\},
	\end{align*}
	where $\partial^\ba =\partial^{a_1}\cdots\partial^{a_p}$ with $\ba=(a_1,\ldots,a_p)^\top \in \mbN_0^p$, and $\|\ba\|_1=\sum_{k=1}^p a_k$.
\end{definition}
\begin{condition}\label{hold}
	Each element of the function $\bg\in\calG$  belongs to the H\"{o}lder class $\calH^{\kappa}(\calX, c_0)$.
\end{condition}

Condition \ref{hold} introduces a basic smoothness condition on the weight function, which is mild. 


Let $R^0_{\rm in}(\bw_{\mathcal{X}},\f^\ast) =n^{-1}\sumi\mathbb{E}_{Y|\bX}[\ell\{Y_i, f_{\bw_{\bX_i}}^\ast(\bX_i)\}]-\delta_{n,\ell}\equiv R_{\rm in}(\bw_{\mathcal{X}},\f^\ast)-\delta_{n,\ell}$, $\f^\ast=(f_1^\ast,\ldots,f_M^\ast)$, 
$\calW_{\mathcal{X}}=\cup_{i=1}^n\calW_{\bX_i}$, $\xi_n=\inf_{\bw_{\mathcal{X}}\in\calW_{\mathcal{X}}}R^0_{\rm in}(\bw_{\mathcal{X}},\f^\ast)$, and $\wt\xi_n=\min\{1,\xi_n\}$.

\begin{condition}\label{conxi4}
	The dimension of $\bX$ and the number of PTMs satisfy the following conditions:
	\begin{align*}
		&M/(a_N \wt\xi_n)=o_p(1),~ M/(n\wt\xi_n)=o_p(1),~ M\wt\xi_n^{-1}p^{\lfloor\kappa\rfloor+\max(\kappa,1)/2}n^{-\frac{\kappa}{(2\gamma+2)p}}=o_p(1),\\
		&\mbox{and}\;\{MSD\log(S)\log(n)\}^{\gamma+1/2}n^{-1/2}\wt\xi_n^{-1}=o_p(1).
	\end{align*}
	where $D-1$ and $S$ are the number of hidden layers and the total number of parameters in the neural network \eqref{w_NN_x}, respectively.
\end{condition}
Condition \ref{conxi4} pertains to the dimensionality of the covariates, the depth and size of the neural network, and the number of PTMs in relation to $n$ and $\xi_n$. 
\begin{Remark}
	Condition \ref{conxi4} involves the term $n^{-\frac{\kappa}{(2\gamma+2)p}}$, which leads to slower convergence when the dimension $p$ is large. In this case, one may apply dimension reduction methods or assume that the support of $\bX$ is concentrated on some neighborhood of a low-dimensional manifold \citep{Chen2022Nonparametric,jiao2023deep}, which can substantially improve the convergence rate.
\end{Remark}
Define $W$ as the maximum width of hidden layers in the neural network \eqref{w_NN_x}. 
We now establish the asymptotic optimality of the proposed estimator with respect to the in-sample risk function. 
\begin{Theorem}\label{opt}
	Under Conditions \ref{convF}--\ref{conxi4}, if $D\asymp n^{1/(4\gamma+4)}\lceil \log_2(8n^{1/(4\gamma+4)})\rceil$ and $W\asymp p^{\lfloor\kappa\rfloor+1}$, we have
	\begin{align}\label{ratioR}
		\frac{R^0_{\rm in}(\wh\bw_{\mathcal{X}},\f^\ast)}{\inf_{\bw_{\mathcal{X}}\in\calW_{\mathcal{X}}}R^0_{\rm in}(\bw_{\mathcal{X}},\f^\ast)}\pover 1,
	\end{align} 
	as $n \rightarrow \infty$.
\end{Theorem}
This theorem shows that our LocalMA estimator is optimal in the sense that its in-sample risk is asymptotically identical to that of the infeasible best model averaging estimator. Unlike previous work on optimal model averaging, our analysis additionally accounts for the approximation error introduced by using the neural network to approximate the true weight function in the H\"{o}lder class.  
The proof of Theorem \ref{opt} is delegated to the Supplementary Material. 

We analyze, through an example, the order of the infimum of the in-sample risks for LocalMA and GlobalMA methods, in order to compare the extent of improvement of LocalMA over GlobalMA. In this example, $\inf_{\bw_{\mathcal{X}}\in\calW_{\mathcal{X}}}R^0_{\rm in}(\bw_{\mathcal{X}},\f^\ast)=O_p(n^{-{1/{10}}})$ and $\inf_{\bw\in\calW}R^0_{\rm in}(\bw,\f^\ast)=\Omega_p(1)$ (for more details, see the Supplementary Material). 
In particular, the above rates imply that the infimum of the in-sample risk achieved by LocalMA vanishes polynomially fast as $n\to\infty$, whereas the infimum of the in-sample risk of GlobalMA remains bounded away from zero, highlighting a strict expressive advantage of input-dependent weights over constant weights.

The next theorem establishes the asymptotic optimality of the proposed estimator in terms of out-of-sample risk, for which we require an additional condition.

\begin{condition}\label{conxistar4}
	Denote $\xi^\ast=\inf_{\bw_{\bX^\ast}\in\calW_{\bX^\ast}}R^0_{\rm out}(\bw_{\bX^\ast})$.
	\begin{align*}
		M\xi^{\ast-1}p^{\lfloor\kappa\rfloor+\max(\kappa,1)/2}n^{-\frac{\kappa}{(2\wt\gamma+2)p}}=o_p(1)
		\mbox{ and}\;\{MSD\log(S)\log(n)\}^{\wt\gamma+1/2}n^{-1/2}\xi^{\ast-1}=o_p(1).
	\end{align*}
\end{condition}
Condition \ref{conxistar4} is similar to Condition \ref{conxi4} and pertains to $p$, $M$, $S$, and $D$ in relation to $n$ and $\xi^\ast$. 
\begin{Theorem}\label{th:opt_R}
	Under Conditions \ref{losssw}--\ref{hold}, and \ref{conxistar4}, if  $D\asymp n^{1/(4\wt\gamma+4)}\lceil \log_2(8n^{1/(4\wt\gamma+4)})\rceil$ and $W\asymp p^{\lfloor\kappa\rfloor+1}$, and $\max_{1\leq m\leq M}\E_\bX\{\wh f_m(\bX)\}^4=O_p(1)$, we have
	\begin{align}
		\frac{R^0_{\rm out}(\wh\bw_{\bX^\ast})}{\inf_{\bw_{\bX^\ast}\in\calW_{\bX^\ast}}R^0_{\rm out}(\bw_{\bX^\ast})}\pover 1,
	\end{align} 
	as $n\rightarrow \infty$.
\end{Theorem}
Compared with Theorem \ref{opt}, Theorem \ref{th:opt_R} is more applicable for model averaging on prediction. 
In the previous example comparing in-sample risks, we can similarly obtain that $\inf_{\bw_{\bX^\ast}\in\calW_{\bX^\ast}}R^0_{\rm out}(\bw_{\bX^\ast})=O_p(n^{-1/{10}})$ and $\inf_{\bw\in\calW}R^0_{\rm out}(\bw)=\Omega_p(1)$. This again highlights the advantage of LocalMA over GlobalMA (for more details, see the Supplementary Material).

The final theorem establishes the consistency of the estimated weights.
Let 
$\calW^\ast_{\bX}=\{{\bw}^\ast_{\bX}:{\bw}^\ast_{\bX}={\rm softmax}\{\bg^\ast(\bX)\},~\bg^{\ast}=\argmin_{\bg\in\calG}R_{\rm in}(\bw_\calX,\f^\ast)\}$ and distance $d({\bw}_{\bX},\calW^\ast_{\bX})=\inf_{{\bw}_{\bX}^{\prime}\in \calW^\ast_{\bX}}\E_\bX\|{\bw}_{\bX}-{\bw}_{\bX}^{\prime}\|_2$.
\begin{Theorem}\label{th:weight}
	Under Conditions \ref{convF}--\ref{conxi4}, if $D\asymp n^{1/(4\gamma+4)}\lceil \log_2(8n^{1/(4\gamma+4)})\rceil$ and $W\asymp p^{\lfloor\kappa\rfloor+1}$, we have
	\begin{align*}
		d(\wh{\bw}_{\bX},\calW^\ast_{\bX})\pover 0,
	\end{align*}
	as $n\to\infty$. 
\end{Theorem}
According to Theorem \ref{th:weight}, the weight function $\wh\bw_\bX$ obtained by \eqref{hatwab} converges to the optimal weight set $\calW^\ast_{\bX}$, in the sense that its distance to $\calW^\ast_{\bX}$ vanishes asymptotically. 
The proof of Theorem \ref{th:weight} is given in the Supplementary Material.

\begin{Remark}
	These theorems indicate that, although the weights are specified in a highly flexible manner, the LocalMA estimator remains manageable and consequently achieves the best possible performance in the weight space.
\end{Remark}

\section{Simulation}\label{sec:simulation}
In this section, we compare the LocalMA method with the GlobalMA method under various settings.

In practice, a PTM uses a large amount of data in the pre-training process and can fit the regression relationship in the data very well. In our simulation, for convenience, we omit the pre-training process of PTMs and directly represent them with the given regression functions. Let $M=3$, and the corresponding regression functions are set as:
\be
&&\wh f_1(\bX)=3X_1^2+2\sin(X_2)+2X_3-X_4+0.5X_5,\n\\
&&\wh f_2(\bX)=-2X_1+X_2-1.5X_3+\cos(X_4)+2X_5^2,\n\\
&&\wh f_3(\bX)=3X_1^2+2\sin(X_2)-2X_3+X_4X_5,\n
\ee
where $\bX=(X_1,\ldots,X_p)^\top$ and $p=5$.

\subsection{Scenario 1}
In this scenario, the data generating process (DGP) of our collected observations is the result of weighting $M$ PTMs: 
\be
Y_i =f_0(\bX_i)+\varepsilon_i = \summ w_m^0(\bX_i)\wh f_m(\bX_i) + \varepsilon_i,
\ee
where $\bX_i\sim\text{N}_p(\mathbf{0}_p,\mathbf{I}_p)$,  $\mathbf{0}_p$ is a $p$-dimensional vector with all elements equal to 0, $\mathbf{I}_p$ is a $p\times p$ identity matrix, $\varepsilon_i\sim\text{N}(0,1)$, 
\be\label{wx_exp}
w_m^0(\bX_i) = \frac{\exp(\btheta_m^\top\bX_i)}{1+\sum_{j=1}^{M-1}\exp(\btheta_j^\top\bX_i)} \text{ for } m=1,\ldots, M-1,
\ee
and $w_M^0(\bX_i)=\{1+\sum_{j=1}^{M-1}\exp(\btheta_j^\top\bX_i)\}^{-1}$.
For the values of the weights in \eqref{wx_exp}, we consider three settings:
\begin{itemize}
	\item[S1.] Let $\theta_{mj}$ be the $j$-th element of $\btheta_m$, equal to $(-1)^{m+j}\sqrt{2m}j^{-2}$ for $m=1,\ldots,M-1$ and $j=1,\ldots,p$.
	
	\item[S2.] For any $i$, let $w_1^0(\bX_i)=\cdots=w_M^0(\bX_i)=1/M$.
	
	\item[S3.] For any $i$, let $w_m^0(\bX_i)=\exp(-m)/\sum_{j=1}^M \exp(-j)$ for $m=1,\ldots,M$.

\end{itemize}

We set $n \in \{100,300,600,900\}$, the training data $\mathcal{D}^{\text{train}}=\{(\bX_i^{\text{train}},Y_i^{\text{train}})\}_{i=1}^n$, $n_{\text{test}}=5000$, and the test data $\mathcal{D}^{\text{test}}=\{(\bX_i^{\text{test}},f_0(\bX_i^{\text{test}}))\}_{i=1}^{n_{\text{test}}}$. 
Our evaluation metrics include the mean squared error (MSE), with $\text{MSE}=n_{\text{test}}^{-1}\sum_{i=1}^{n_{\text{test}}}\{f_0(\bX_i^{\text{test}})-\wh Y_i^{\text{test}}\}^2$, to evaluate predictive performance, and $C_1(\bw)=n_{\text{test}}^{-1}\sum_{i=1}^{n_{\text{test}}}\left\|\wh \bw_{\bX_i}-\bw^0_{\bX_i}\right\|_1$ to
evaluate weight consistency, where $\wh Y_i^{\text{test}}$ is the prediction based on a certain method, $\bw^0_{\bX_i}=(w_1^0(\bX_i),\ldots,w_M^0(\bX_i))^\top$, and $\|\cdot\|_1$ is the $L_1$ norm.
The methods involved in the comparison include: 
\begin{itemize}
	\item Our localized model averaging method (LocalMA). In \eqref{w_NN_x}, we set $D=2$, $p_2=16$, and use Adam optimization with a learning rate of 0.01. The objective of the optimization is \eqref{LossNN}, and the final parameter estimates are obtained after 800 iterations.
	
	\item Global weighting (GW): $\wh f_{\wh\bw}(\bX^\ast)=\summ \wh w_m \wh f_m(\bX^\ast)$, where $\wh\bw=(\wh w_1,\ldots, \wh w_M)^\top =\argmin_{\bw\in\calW}\sum_{i=1}^n\{Y_i-\wh f_\bw(\bX_i)\}^2$.
	
	\item Equal weight model averaging (EWMA): $\wh f_{\bw_{\mathrm{EW}}}(\bX^\ast)=M^{-1}\summ \wh f_m(\bX^\ast)$, where $\bw_{\mathrm{EW}}=(1/M,\ldots, 1/M)^\top$.
	
\end{itemize}
The observations are repeatedly generated 500 times, and the MSE for each method and $C_1(\bw)$ are obtained each time. The mean values of these 500 MSEs and $C_1(\bw)$ under settings S1-S3 are reported in Tables \ref{tab:MSPE_2NN} and \ref{tab:weight}, respectively. 

In Table \ref{tab:MSPE_2NN}, when both the GW and EWMA methods are misspecified (setting S1), the LocalMA method demonstrates a substantial improvement over these two methods. When the GW method is correctly specified (settings S2 and S3), although the LocalMA method does not outperform GW, the predictive performances of the two methods become very close as $n$ increases. 
In summary, the LocalMA method is significantly better than GW in certain settings, while in cases where the GW method dominates, the predictive performance of LocalMA gradually approaches that of GW as $n$ increases. 
In Table \ref{tab:weight}, $C_1(\bw)$ decreases with increasing $n$ in all settings (S1-S3). This phenomenon indicates that our estimated weights converge to the true weights.

\begin{table}[htbp]
	\setlength{\tabcolsep}{7mm}
	\centering
	\caption{Comparison of MSE for different methods.}\label{tab:MSPE_2NN}
	\begin{tabular}{ccccc}
		\hline
		Setting &$n$   & \multicolumn{1}{c}{LocalMA} & \multicolumn{1}{c}{GW} &EWMA  \\
		\hline
		S1 &100   & \textbf{0.36} (0.15)  & 6.34 (0.64) & 6.03 (0.28) \\
		S1 &300 &\textbf{0.10} (0.03)   & 6.05 (0.38) & 6.04 (0.28)  \\
		S1 &600 & \textbf{0.05} (0.02)  & 5.97 (0.33) & 6.03  (0.29) \\
		S1 &900 & \textbf{0.04}  (0.01)   & 5.95 (0.31) & 6.04 (0.28) \\\hline
		S2 &100   & 0.38 (0.15)  &  0.02 (0.02)  & \textbf{0.00} (0.00) \\
		S2 &300 &  0.10 (0.04) & 0.01 (0.01)  & \textbf{0.00} (0.00)  \\
		S2 &600 & 0.05 (0.01) & 0.003 (0.003) &  \textbf{0.00} (0.00)\\
		S2 &900 & 0.03 (0.01)  &  0.002 (0.002) &  \textbf{0.00} (0.00) \\\hline
		S3 &100   & 0.31 (0.13) & \textbf{0.02}  (0.02) &2.08  (0.04) \\
		S3 &300 & 0.09 (0.03) & \textbf{0.01} (0.01)  & 2.08 (0.04)  \\
		S3 &600 & 0.04 (0.01)  & \textbf{0.003} (0.003)  & 2.08 (0.04)  \\
		S3 &900 &  0.03  (0.01)   &  \textbf{0.002} (0.002) &  2.08 (0.05) \\\hline
	\end{tabular}%
	\begin{tablenotes}
		\footnotesize
		\item[1] Note: The smallest MSE in each row is shown in bold and standard errors are in parentheses.
	\end{tablenotes}
\end{table}%

\begin{table}[htbp]
	\setlength{\tabcolsep}{6.5mm}
	\centering
	\caption{Average of $C_1(\bw)$ over 500 replications.}\label{tab:weight}
	\begin{tabular}{ccccc}
		\hline
		Setting & $n$=100 & $n$=300 &$n$=600 &$n$=900 \\
		\hline
		S1   & 0.20 (0.04)  & 0.12 (0.02) & 0.09 (0.02) & 0.08 (0.01) \\
		S2   &  0.20 (0.04) & 0.10 (0.02) &0.07 (0.01) &0.06 (0.01) \\
		S3   & 0.17 (0.04)  & 0.10 (0.02) & 0.07 (0.01) & 0.06 (0.01) \\
		\hline
	\end{tabular}%
	\begin{tablenotes}
		\footnotesize
		\item[1] Note: Standard errors are in parentheses.
	\end{tablenotes}
\end{table}%

\begin{figure}[htbp]
	\centering
	\subfigure[]{
		\includegraphics[width=7cm]{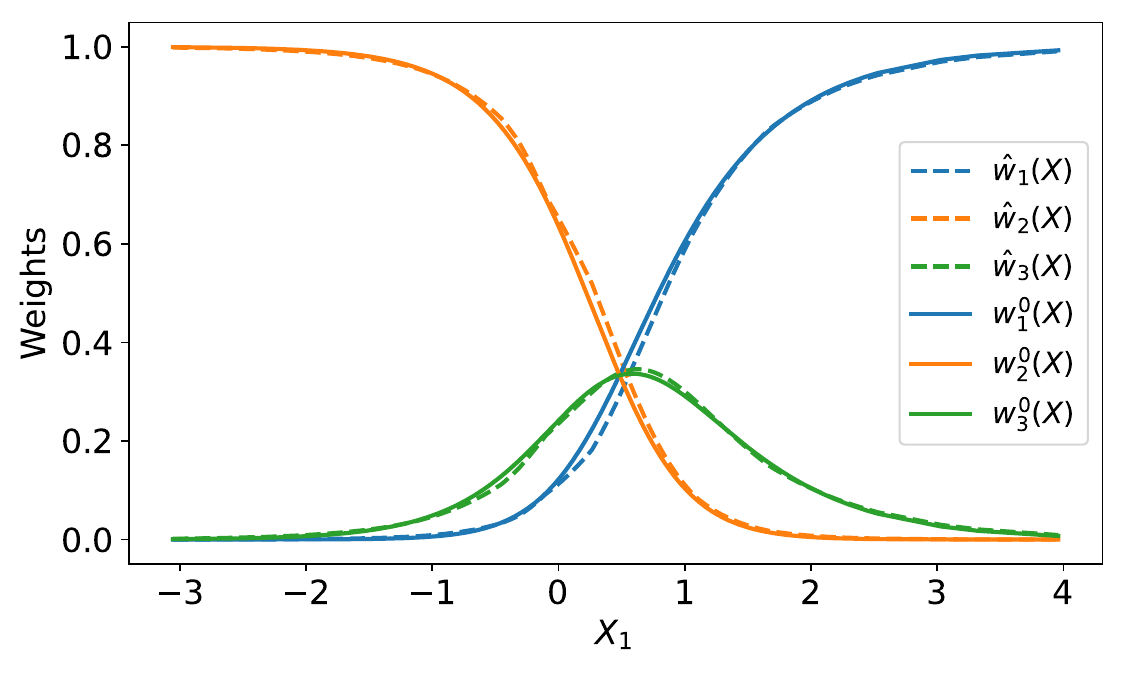}
	}
	\subfigure[]{
		\includegraphics[width=7cm]{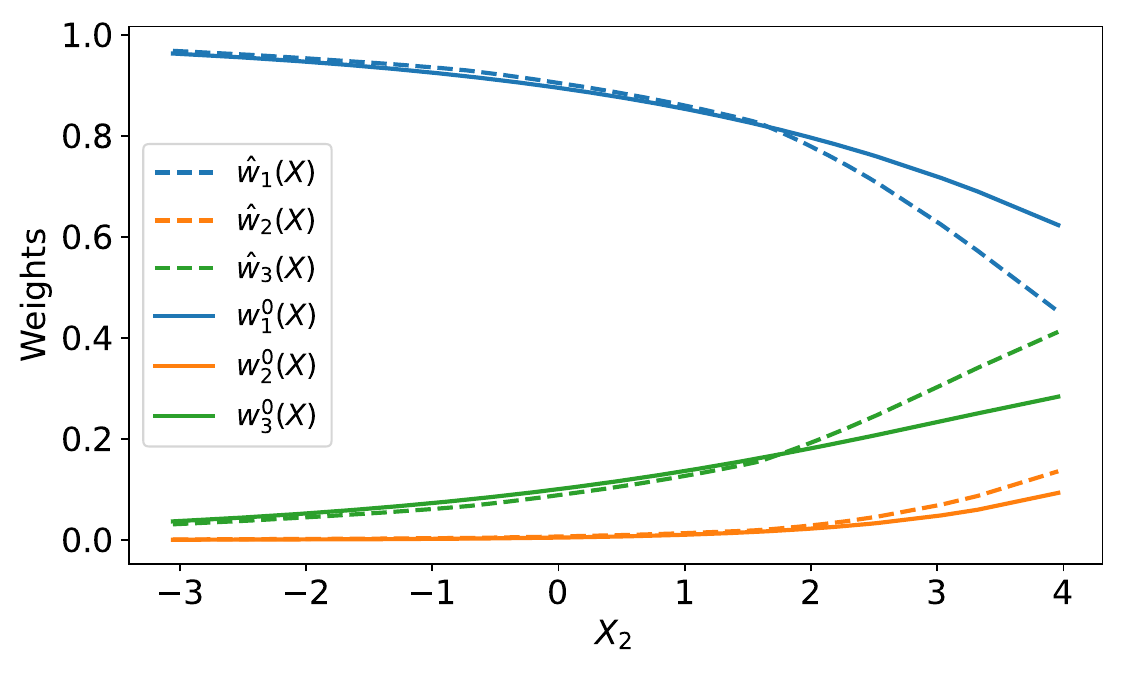}
	}
	\subfigure[]{
		\includegraphics[width=7cm]{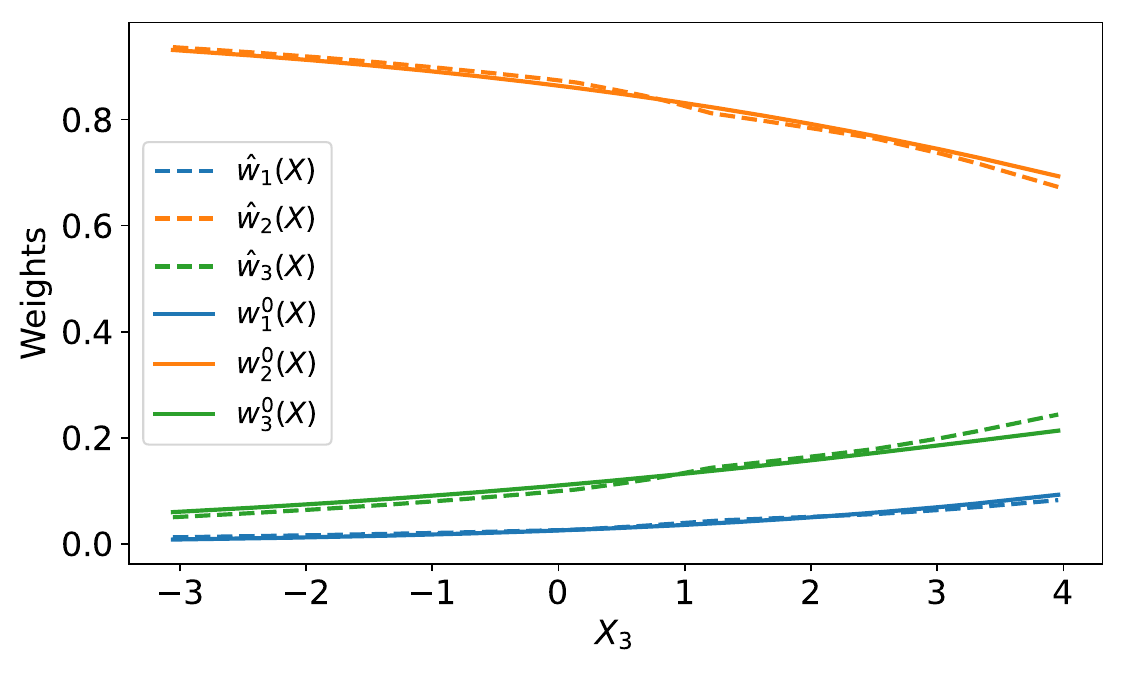}
	}
	\subfigure[]{
		\includegraphics[width=7cm]{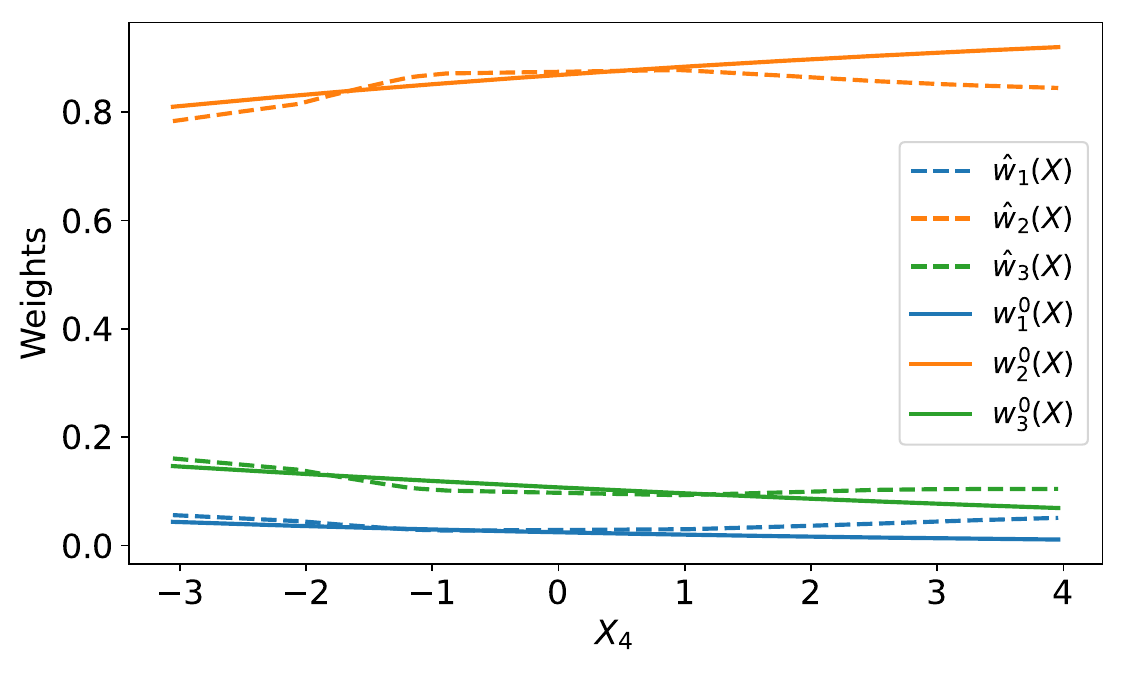}
	}
	\subfigure[]{
		\includegraphics[width=7cm]{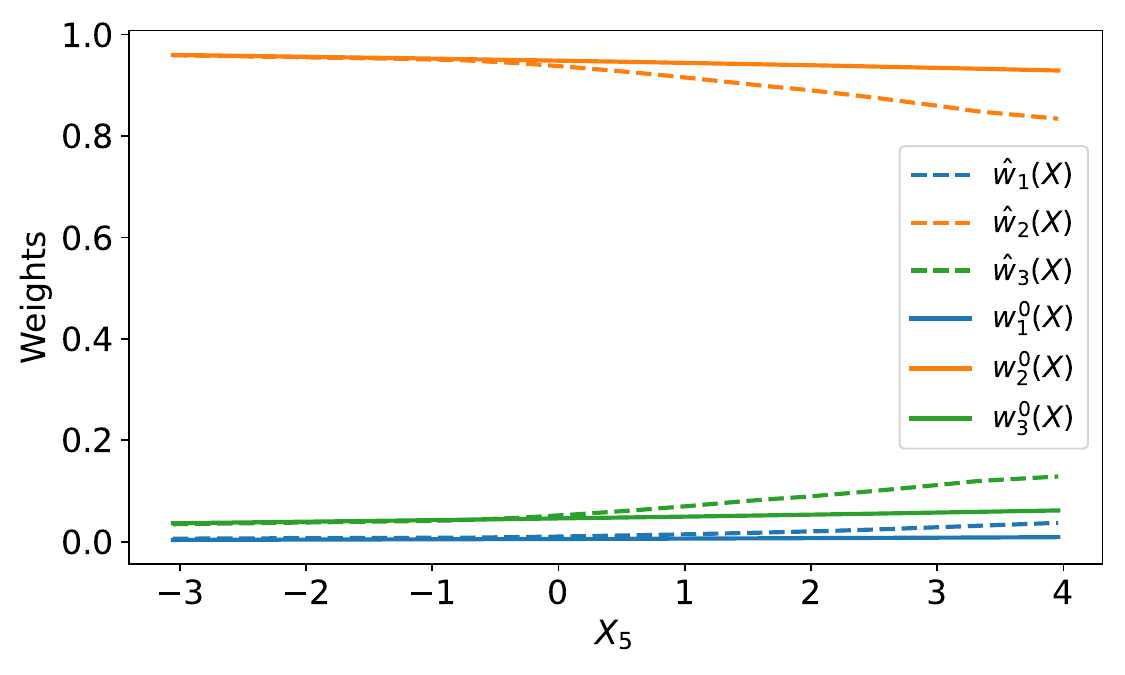}
	}
	\caption{The true weight functions and the estimated weight functions in setting S1.} \label{fig:LMA_weight}
\end{figure}

To intuitively demonstrate the effectiveness of weight estimation, we plot the estimated weight functions alongside the true weight functions for each covariate in setting S1. When plotting the weight functions with respect to the $j$-th covariate $X_j$, we fix all other covariates and compute the values of the weight functions over $X_j$. As shown in Figure \ref{fig:LMA_weight}, the solid lines represent the true weight functions, while the dashed lines represent the estimated weight functions. Overall, the estimation of the weight functions performs reasonably well.

\subsection{Scenario 2}
In this scenario, we consider the case where DGP is not given by a weighted combination of the PTMs, that is,
\be
Y_i=f_0(\bX_i)+\varepsilon_i=3X_{i1}^2+2\sin(X_{i2})-2X_{i3}+\cos(X_{i4})+2X_{i5}^2+\varepsilon_i,\quad i=1,\ldots,n,\n
\ee
where $\bX_i=(X_{i1},\ldots,X_{ip})^\top\sim\text{N}_p(\mathbf{0}_p,\mathbf{I}_p)$, $\varepsilon_i\sim\text{N}(0,1)$, and $p=5$. The other settings are the same as those in Scenario 1. The observations are repeatedly generated 500 times, and the mean values of 500 MSEs are reported in Table \ref{tab:MSPE_miss}. It can be seen that the predictive performance of the LocalMA method is significantly better than that of the other methods.

\begin{table}[htbp]
	\setlength{\tabcolsep}{8mm}
	\centering
	\caption{Comparison of MSE for different methods.}\label{tab:MSPE_miss}
	\begin{tabular}{cccc}
		\hline
		$n$   & \multicolumn{1}{c}{LocalMA} & \multicolumn{1}{c}{GW} &EWMA  \\
		\hline
		100   & \textbf{5.61} (1.27) & 14.50 (0.96) & 16.11 (0.44) \\
		300 &\textbf{4.00}  (0.31)  & 14.09 (0.53) & 16.10 (0.43)  \\
		600 & \textbf{3.74} (0.24)  & 13.98 (0.47)  & 16.10  (0.44) \\
		900 & \textbf{3.66}  (0.23)   & 13.95 (0.46) & 16.11 (0.45) \\
		\hline
	\end{tabular}%
	\begin{tablenotes}
		\footnotesize
		\item[1] Note: The smallest MSE in each row is shown in bold and standard errors are in parentheses.
	\end{tablenotes}
\end{table}%

\section{Empirical evaluation}\label{sec:dataset}
In this section, we use three datasets to evaluate the practical performance of the LocalMA method: the China rental housing dataset for a regression task, the multilingual toxicity detection dataset for a text classification task, and the CIFAR-10 dataset for an image classification task.

\subsection{China rental housing dataset}
We use the rental housing data from all districts of Beijing and Shanghai in this dataset\footnote{\url{http://www.idatascience.cn/dataset}}, where the response variable $Y$ is the natural logarithm of the monthly rent, and the covariates $\bX$ are the number of rooms, the number of restrooms, the number of living rooms, total area, have or not a bed, have or not a wardrobe, have or not an air conditioner, have or not a fuel gas, total floor, the number of schools within 3 kilometers, as recommended by \cite{hu2023optimal}. The goal is to predict monthly rent using 10 covariates. 
To mitigate the impact of extreme observations, we excluded those with monthly rents above the 99th percentile. 
The number of observations for all districts in Beijing and Shanghai is 4,654 and 5,432, respectively. The number of observations for each district is shown in Table \ref{tab:number_CRD}.
\begin{table}[htbp]
	\setlength{\tabcolsep}{0.8mm}
	\centering
	\caption{The number of observations in each district of Beijing and Shanghai.}\label{tab:number_CRD}
	\begin{tabular}{ccccccc}
		\hline
		Changping & Chaoyang & Dongcheng &Haidian &Shijingshan &Fengtai &Daxing \\
		310   & 1215  & 289  & 518 & 269 &345 &291 \\\hline
		Xicheng &Tongzhou &Fangshan &Shunyi &Huairou &Mentougou &Jiading\\
		299 &223 &249 &220 &162 &264 &302 \\
		\hline
		Fengxian   &Baoshan  &Xuhui   &Putuo   &Yangpu   & Pudong  & Hongkou  \\
		204  &365   &563   & 416  & 316  & 1311  & 207  \\\hline
		Changning &Minhang  &Jingan  &Huangpu &&&\\
		431  &584  &347  &386 &&&\\\hline
	\end{tabular}%
\end{table}

Our method requires obtaining PTMs and training the weight function. Taking Beijing as an example, we split the observations from Beijing into two parts according to proportion $1-\rho$ vs $\rho$: $\mathcal{D}_1$ is used to build the PTMs, and $\mathcal{D}_2$ is used to train the weight function and to test the performance of all methods. We use neural networks (a feedforward neural network with one hidden layer of 32 neurons, ReLU activation, and a linear output layer for scalar prediction) or random forests (with 100 trees) built separately on the observations from each district in $\mathcal{D}_1$ as PTMs, i.e., $M=13$. 
In $\mathcal{D}_2$, after removing half of the observations from Haidian and Chaoyang districts, we evenly split the remaining data into a training set $\mathcal{D}_2^{\train}$ and a testing set $\mathcal{D}_2^{\test}$. 
The neural network (a feedforward neural network with one hidden layer of 32 neurons and ReLU activation, followed by an output layer with $M$ neurons and a softmax transformation) trained on $\mathcal{D}_2^{\train}$ is used to estimate the weight function for combining the PTMs, and $\mathcal{D}_2^{\test}$ is used to evaluate the performance of all methods.

\begin{figure}[htbp]
	\centering
	\subfigure[PTMs and DirectModel are neural networks]{
		\includegraphics[width=7cm]{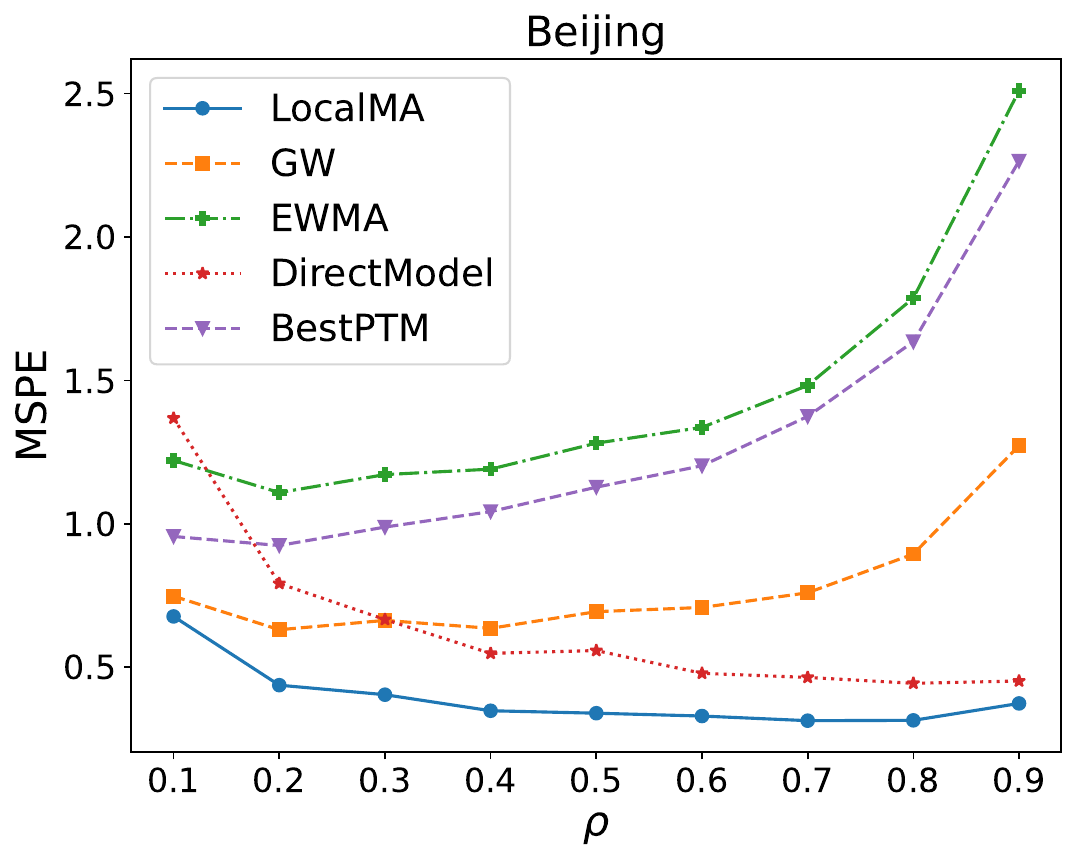}
	}
	\subfigure[PTMs and DirectModel are random forests]{
		\includegraphics[width=7.2cm]{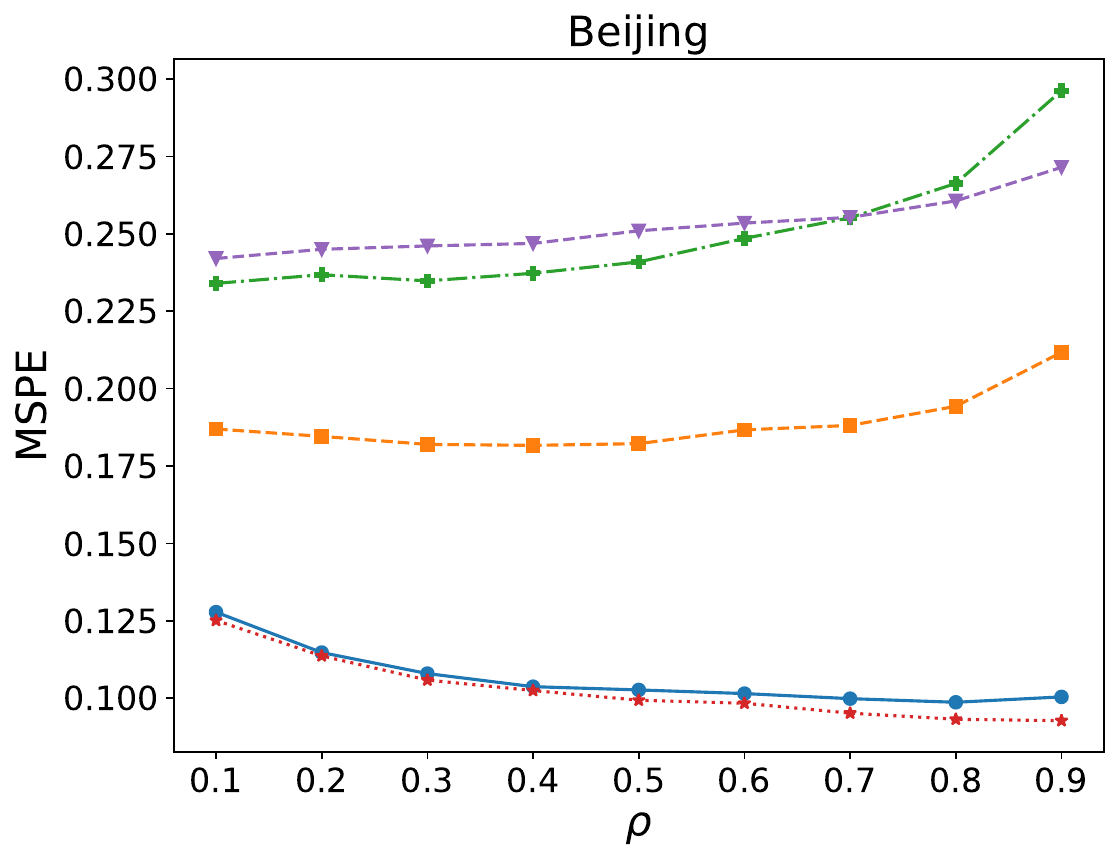}
	}
	\subfigure[PTMs and DirectModel are neural networks]{
		\includegraphics[width=7cm]{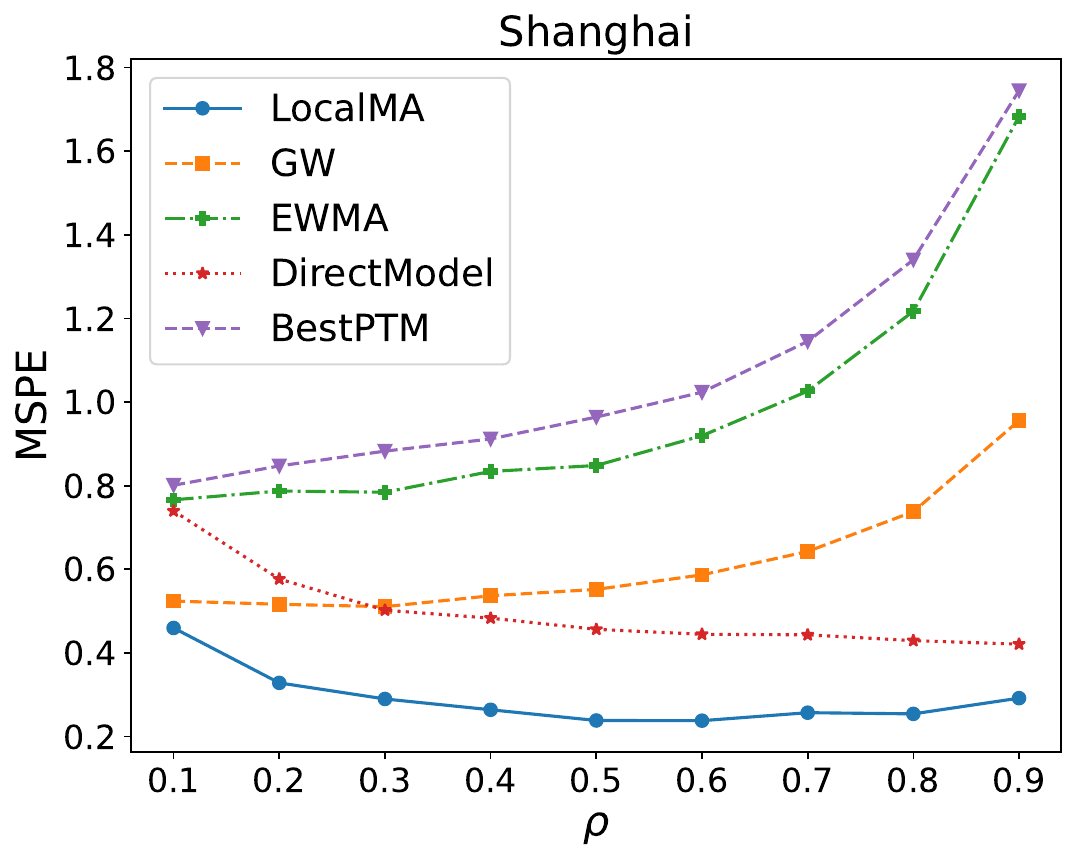}
	}
	\subfigure[PTMs and DirectModel are random forests]{
		\includegraphics[width=7.2cm]{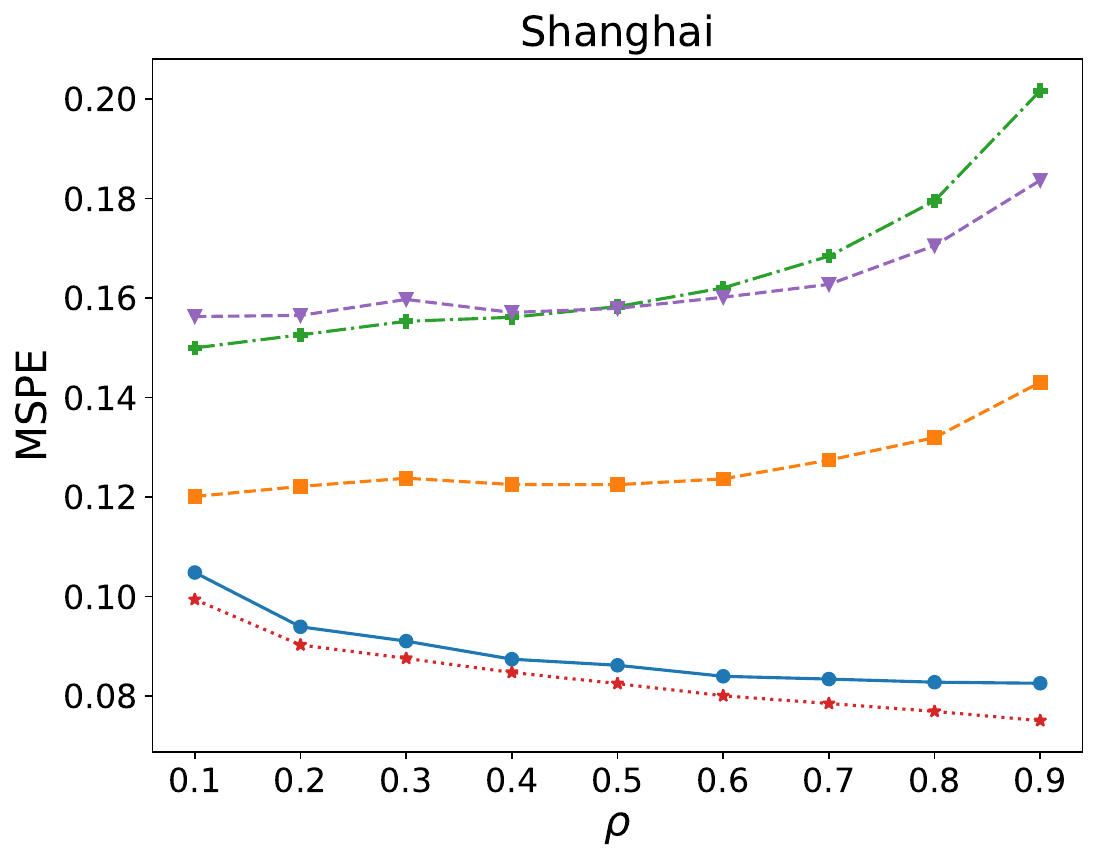}
	}
	\caption{Comparison of MSPE for different methods.} \label{fig:rental}
\end{figure}

On the testing set $\mathcal{D}_2^{\test}$, we use mean squared prediction error (MSPE) to compare the following methods: our method (LocalMA), the GW method, the EWMA method, the method that directly trains a neural network or a random forest on $\mathcal{D}_2^{\train}$ to predict $Y$ (DirectModel), and the best-performing PTM on $\mathcal{D}_2^{\test}$ (BestPTM). The evaluation metric is ${\rm MSPE}=n^{-1}_{\rm test}\sum_{i=1}^{n_{\rm test}}(Y_i^{\rm test}-\wh Y_i^{\rm test})^2$, where $n_{\rm test}$ is the number of observations in the testing set $\mathcal{D}_2^{\test}$, $Y_i^{\rm test}$ is the response, and $\wh Y_i^{\rm test}$ is the prediction based on each method. 
Let $\rho\in\{0.1,0.2,\ldots,0.9\}$. For each $\rho$, we repeat the experiment 100 times, and the mean values over 100 simulation results are displayed in Figure \ref{fig:rental}. The following key findings emerge from Figure \ref{fig:rental}: (1) As $\rho$ increases, the MSPE of the DirectModel method tends to decrease, because more training data usually leads to better model performance. (2) The MSPEs of the BestPTM and EWMA methods increase as $\rho$ increases, because less data is available for building the PTMs. (3) For the remaining two methods that require weight estimation (LocalMA, GW), a larger $\rho$ means more observations are available for estimating the weights, but fewer are left for building the PTMs. This creates a trade-off for the LocalMA and GW methods, so their MSPEs are generally non-monotonic. (4) It can be observed that the gap in MSPE between LocalMA and GW/EWMA becomes larger as $\rho$ increases. This indicates that the three methods differ in how effectively they leverage the predictions of PTMs, with the LocalMA method demonstrating the most efficient utilization. (5) When the PTMs and the DirectModel method use neural networks, the LocalMA method always achieves the best predictive performance; when the PTMs and the DirectModel method use random forests, the DirectModel method performs best, but the LocalMA method is very close to it.

\subsection{Multilingual toxicity detection dataset}
We consider the multilingual toxicity detection dataset\footnote{\url{https://huggingface.co/datasets/textdetox/multilingual_toxicity_dataset}} provided by \cite{dementieva2024overview} and use the TinyBERT\footnote{\url{https://huggingface.co/huawei-noah/TinyBERT_General_4L_312D}} model provided by \cite{jiao2019tinybert} as the basis for predicting whether a sentence in a given language is toxic. We select six languages from this dataset: Chinese (5,000 observations), English (5,000), French (5,000), Ukrainian (5,000), Hebrew (2,011), and Hindi (4,363). The covariate is a sentence in a certain language, and the response is binary. If a sentence is toxic, the corresponding response takes the value 1; otherwise, it is 0. Therefore, this is a text classification task. 

We refer to Hebrew and Hindi, which have fewer observations, as minority languages, while the other languages are considered non-minority languages. The goal is to leverage data from the non-minority languages to improve the accuracy of toxic text classification in the minority languages.
The minority languages are randomly split into a training set $\mathcal{D}^{\text{train}}$ and a test set $\mathcal{D}^{\text{test}}$ according to the ratio $(1-\rho):\rho$. 
We use the TinyBERT model built separately on the observations from each non-minority language as PTMs, i.e., $M=4$. The TinyBERT model trained on the training set $\mathcal{D}^{\text{train}}$ is used to estimate the weight function for combining the PTMs. 

\begin{figure}[htbp]
	\centering
	\includegraphics[width=10cm]{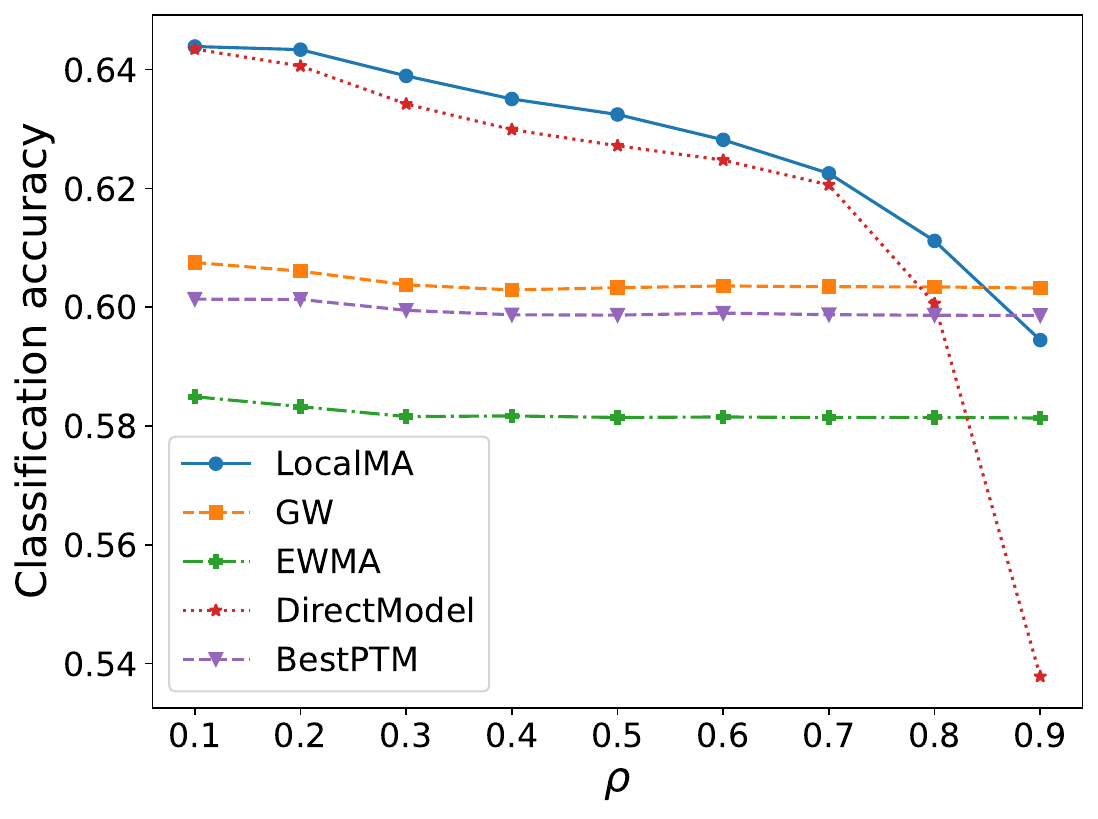}
	\caption{Comparison of the classification accuracy for different methods.} \label{fig:NLP}
\end{figure}

On the testing set $\mathcal{D}^{\text{test}}$, we use classification accuracy to compare the following methods: our method (LocalMA), the GW method, the EWMA method, the method that directly trains the TinyBERT model on the training set $\mathcal{D}^{\text{train}}$ to predict $Y$ (DirectModel), and the best-performing PTM on the testing set $\mathcal{D}^{\text{test}}$ (BestPTM). Let $\rho\in\{0.1,0.2,\ldots,0.9\}$. For each $\rho$, we repeat the data splitting 50 times, and the mean values over these 50 repetitions are presented in Figure \ref{fig:NLP}. 
There are four main findings from Figure \ref{fig:NLP}: (1) As $\rho$ increases, the number of observations in the training set $\mathcal{D}^{\text{train}}$ decreases, leading to lower classification accuracy for the LocalMA method and the DirectModel method.
(2) The classification accuracy of the BestPTM method and the EWMA method hardly changes with $\rho$, since the PTMs are unaffected by $\rho$.
(3) The classification accuracy of the GW method hardly changes with $\rho$, possibly because the global weights cannot improve its performance.
(4) In most cases, the LocalMA method achieves the highest classification accuracy.

\subsection{CIFAR-10 dataset}\label{sec:cifar10}
The CIFAR-10 dataset \citep{krizhevsky2009learning} is a widely used benchmark for image classification tasks in computer vision research. It consists of 60,000 32$\times$32 color images, divided into ten categories (airplane, automobile, bird, cat, deer, dog, frog, horse, ship, truck), with 6,000 images per category. The dataset is split into 50,000 training images in $\mathcal{D}^{\text{train}}$ and 10,000 testing images in $\mathcal{D}^{\text{test}}$. 
A series of data augmentation techniques is applied to the training images: (1) random horizontal flip; (2) random rotation within $\pm10$ degrees; and (3) random adjustments to brightness and contrast (both with a factor of 0.2). 

We consider the following scenario: four companies each own one of the following models: ResNet18 (11.7M), DenseNet121 (8M), GoogLeNet (6.6M), and EfficientNet$\underline{~}$B0 (5.3M)\footnote{\url{https://docs.pytorch.org/vision/stable/models.html}}  pre-trained on the ImageNet-1K dataset. 
Because the ImageNet-1K dataset has 1,000 categories while the CIFAR-10 dataset has only 10, these companies fine-tune their own models by replacing the final layer's 1,000 neurons with 10 neurons and training only the parameters of that final layer on their own collected data, leaving the other parameters unchanged. 
The data collected by each company differ; for example, the company using model ResNet18 collects all training images $\mathcal{D}^{\text{train}}$ labeled airplane, automobile, bird, and cat, together with 50 randomly selected images from the remaining labels. The company using DenseNet121 collects all training images $\mathcal{D}^{\text{train}}$ labeled bird, cat, deer, and dog, plus 50 randomly selected images from the other labels.
The company using GoogLeNet collects all training images $\mathcal{D}^{\text{train}}$ labeled deer, dog, frog, and horse, plus 50 randomly selected images from the other labels.
The company using EfficientNet$\underline{~}$B0 collects all training images $\mathcal{D}^{\text{train}}$ labeled frog, horse, ship, and truck, plus 50 randomly selected images from the other labels.

Assume that we randomly collect $50,000\times\rho$ images $\mathcal{D}_\rho^{\text{train}}$ from the training images $\mathcal{D}^{\text{train}}$, we use each company's fine-tuned models as PTMs ($M=4$), and the convolutional neural network (see Supplementary Material for more implementation details) trained on the collected images $\mathcal{D}_\rho^{\text{train}}$ is used to estimate the weight function for combining the PTMs.
On the testing images $\mathcal{D}^{\text{test}}$, we use classification accuracy to compare the following methods: our method (LocalMA), the GW method, the EWMA method, and four PTMs. Let $\rho\in\{0.1,0.2,\ldots,1.0\}$. For each $\rho$, we repeat the random collection 50 times, and the mean values over these 50 repetitions for the LocalMA and GW methods are presented in Table \ref{tab:CIFAR-10}. Regardless of the value of $\rho$, the classification accuracies of the four PTMs (ResNet18, DenseNet121, GoogLeNet, EfficientNet$\underline{~}$B0) and the EWMA method are 35.04\%, 30.88\%, 33.25\%, 37.46\%, and 65.05\%, respectively.

There are four main findings from the results: (1) As the number of collected images $\mathcal{D}_\rho^{\text{train}}$ increases (i.e., as $\rho$ increases), the classification accuracy of the LocalMA method improves. (2) The classification accuracy of the GW method remains almost unchanged with respect to $\rho$, possibly because global weights are difficult to use to enhance predictive performance. (3) The performances of the other methods are invariant to $\rho$, since the PTMs are directly applied to the testing images $\mathcal{D}^{\text{test}}$. (4) The LocalMA method consistently achieves the highest classification accuracy, indicating that locally weighting the PTMs is reasonable.

\begin{table}[htbp]
	\setlength{\tabcolsep}{4mm}
	\centering
	\caption{Comparison of the classification accuracy for LocalMA and GW methods.}\label{tab:CIFAR-10}
	\begin{tabular}{cccccccc}
		\hline
		&$\rho=0.1$ &$\rho=0.2$ &$\rho=0.3$ &$\rho=0.4$ &$\rho=0.5$ \\
		LocalMA  & 66.1434\% & 67.5148\% & 67.2756\% & 67.3796\% & 67.3982\%  \\
		GW  & 64.8724\% & 64.8738\% & 64.8732\% & 64.8720\% & 64.8714\%
		\\\hline
		&$\rho=0.6$ &$\rho=0.7$ &$\rho=0.8$ &$\rho=0.9$ &$\rho=1.0$ \\
		LocalMA  &68.0452\% & 68.5384\% & 69.0042\% & 69.4404\% & 69.9746\%  \\
		GW  &64.8738\% & 64.8720\% & 64.8722\% & 64.8716\% & 64.8722\% 
		\\\hline
	\end{tabular}%
\end{table}%

\addtolength{\textheight}{-.2in}%

\section{Concluding remarks}\label{sec:concluding}
Given that PTMs provided by different companies often exhibit varying performances across tasks, assigning input-dependent weights is crucial when combining these PTMs. This allows the method to adaptively assign higher weights to PTMs that perform better in specific regions of the input space, making our proposed LocalMA method more flexible than GlobalMA. 
In a general framework that allows for unbounded loss functions, we establish the asymptotic optimality of both in-sample and out-of-sample risks, as well as the consistency of the estimated weights. Despite the flexibility of the weights, our estimators remain well-behaved and achieve the best possible performance over the weight space. In addition, the proposed method achieves favorable performance across three real-world datasets. 


In future work, we plan to establish oracle inequalities for MoE and provide a theoretical understanding of its performance, and further investigate the fundamental reasons behind its remarkable success in large language models. Such analysis would help reveal the underlying mechanisms that explain the effectiveness of MoE and provide a stronger theoretical foundation for its practical applications.



\section*{Supplementary material}
The online supplementary material contains some examples and the proofs of the theorems.

\bibliography{RefList.bib}

\end{document}